\documentclass[english,aps,prc,0superscriptaddress,longbibliography,notitlepage]{revtex4-1}
\usepackage{amsmath,amssymb,multirow,bm}
\usepackage{soul}
\usepackage{epsfig}
\usepackage{graphicx}
\usepackage{amsmath}
\usepackage{lipsum}
\usepackage{color}
\makeatother

\usepackage{babel}
\usepackage[colorlinks,linkcolor=blue,anchorcolor=blue,citecolor=blue,urlcolor=blue,filecolor=black]{hyperref}

\newcommand{\beq}{\begin{equation}}
\newcommand{\eeq}{\end{equation}}
\newcommand{\bea}{\begin{eqnarray}}
\newcommand{\eea}{\end{eqnarray}}
\usepackage{graphicx}
\preprint{}

\begin{document}

\title{Photonuclear Tomography in Ultraperipheral Heavy-Ion Collisions}

\author{J. D. Baker}
\email{jbaker51@leomail.etamu.edu}
\affiliation{Department of Physics and Astronomy, East Texas A\&M University, Commerce, TX 75428, USA}

\author{C. A. Bertulani}
\email{carlos.bertulani@etamu.edu}
\affiliation{Department of Physics and Astronomy, East Texas A\&M University, Commerce, TX 75428, USA}
\affiliation{ExtreMe Matter Institute EMMI, GSI Helmholtzzentrum für Schwerionenforschung GmbH, Planckstrasse 1, 64291 Darmstadt, Germany}
\affiliation{Institut f\"ur Kernphysik, Technische Universit\"at Darmstadt, Schlossgartenstr. 9, 64289 Darmstadt, Germany}

\author{Victor P. Gon\c{c}alves}
\email{barros@ufpel.edu.br}
\affiliation{Institute of Physics and Mathematics, Federal University of Pelotas (UFPel), 
  Postal Code 354,  96010-900, Pelotas, RS, Brazil}

\date{\today}

\begin{abstract}
We present a theoretical investigation of photonuclear tomography as a novel technique for probing the internal structure of nuclei. In this approach, ultraperipheral heavy-ion collisions (UPCs) serve as a source of intense fluxes of virtual photons, which induce coherent production of vector mesons. By analyzing the probabilities and cross sections of these photon-induced processes, we propose a methodology for reconstructing the spatial distribution of nucleons within the nucleus. Our framework provides a systematic way to access information on the  nuclear geometry probed in UPCs, offering new opportunities for studies of nuclear structure using particle production as a probe. Numerical calculations for selected examples illustrate the feasibility and potential of this method.

\end{abstract}

\maketitle

\section{Introduction}
Ultraperipheral heavy-ion collisions (UPCs) provide an ideal environment to search for exotic hadronic states due to their inherently low event multiplicity and the possibility of 
efficient background suppression~\cite{ISI:A1988P121800001,annurev.nucl.55.090704.151526}. In such collisions, the impact parameter exceeds the sum of the radii of the 
colliding ions, allowing them to interact predominantly via electromagnetic processes while remaining elastically scattered. This results in the absence of additional calorimetric activity and the presence of large rapidity gaps between the produced particles and the outgoing beams - features that facilitate effective background rejection and allow us to perform a more detailed investigation of the nuclear structure~\cite{annurev.nucl.55.090704.151526}.

Throughout this study, we concentrate on particle production in UPCs, with particular emphasis on vector meson production. Owing to the fact that photons possess the same quantum numbers as vector mesons, a high-energy photon beam effectively behaves as a beam of quark-antiquark pairs. The flavor content of these pairs depends on the specific vector meson produced: light quarks correspond to $\rho$ or $\omega$ mesons, strange quarks to $\phi$ mesons, and charm quarks to $J/\psi$ mesons. Our focus lies on processes in which a quasi-real photon emitted by one of the ions fluctuates into a vector meson state $V$ with invariant mass $M_V$, subsequently interacting with the other incoming ion at the interaction point.

The cross section for such processes factorizes into two contributions: the photon flux (or photon luminosity) generated by one ion and the photonuclear production cross section $\sigma_{\gamma A}$ for the creation of $V$ via photon-nucleus interactions~\cite{ISI:A1988P121800001,PhysRevC.60.014903}. Instead of adopting the traditional momentum-space description common in quantum field theory, we employ a coordinate-space formulation, which provides additional insights into the spatial regions within the nucleus where the photon interacts and induces particle production.

The conventional calculation of photonuclear processes in UPCs involves the convolution of the equivalent photon spectrum generated by the ion with photon-induced reaction cross sections. Typically, this procedure is performed in momentum space because of the simplicity of the photon propagator in this representation. However, the problem also lends itself naturally to a coordinate-space analysis, which facilitates the exploration of the spatial dependence of photon interactions within the nucleus. This approach also enables an assessment of the validity of the equivalent photon approximation, which assumes first-order photon exchange~\cite{ISI:A1988P121800001}. In this work, we examine the applicability of this approximation by considering the high-energy tail of the photon spectrum (in the nucleus rest frame), focusing on photon energies of approximately 1~GeV and above.

Our analysis focuses on the production of $J/\psi$ and $\rho^0$ mesons, which have been extensively studied in UPCs at the LHC over the past two decades. In particular, we concentrate on the elastic diffractive process, in which the nucleus target remains intact after the interaction, aiming to highlight the role of nuclear geometry and the localization of photonuclear interactions in UPCs. We demonstrate that vector meson production in UPCs at the Large Hadron Collider (LHC) at CERN may serve as a tool for probing nuclear mass and charge distributions, with sensitivity depending on the mass of the produced meson. Specifically, we consider collisions at a nucleon-pair center-of-mass energy of $\sqrt{s_{NN}} = 5.5$~TeV, corresponding to lead-lead (Pb+Pb) collisions, which are currently under experimental investigation at the LHC.

We further explore the possibility of experimentally accessing nuclear geometry details through vector meson production. Accurate measurements of the neutron-skin thickness in nuclei provide important experimental constraints on the symmetry energy near the nuclear saturation density $n_0 = 0.17$~fm$^{-3}$ and on the equation of state (EoS) of asymmetric nuclear matter~\cite{typel:2001:PRC,centelles:2009:PRL,TamiPRL.107.062502,PhysRevLett.108.112502,AumannPRL119}. The EoS describes the energy per nucleon, or the pressure, as a function of density $n$ and is crucial for astrophysical studies, including neutron star mass-radius relationships and gravitational wave signatures from neutron star mergers. The symmetry energy, $E_{\mathrm{sym}}(n)$, and particularly its slope parameter, $L = 3 n_0 \, dE_{\mathrm{sym}}(n) / d\rho|_{n_0}$, remain poorly constrained and exhibit significant model dependence across different nucleon-nucleon interactions~\cite{Lattimer:2001,2005PhR...411..325S,Lattimer:2012}.

The neutron skin of a nucleus is defined as the difference between the root-mean-square (rms) radii of the neutron and proton density distributions, $R_{np}=\sqrt{\left< r_n\right>^2 - \left< r_p\right>^2} $~\cite{PhysRevLett.106.252501,HorowitzPRL.86.5647}. While electron scattering experiments~\cite{RevModPhys.28.214} and isotopic shift measurements~\cite{PhysRevA.83.012516} provide reliable information on charge distributions, extracting neutron or matter distributions remains challenging due to the dependence of hadronic probes on nuclear reaction models, leading to significant uncertainties~\cite{AumannPRL119}. Recent studies using the parity-violating component of electron-nucleus interactions have also produced inconsistent results~\cite{PhysRevLett.108.112502,PhysRevLett.126.172502}.

Alternative methods have been proposed for probing neutron skins, such as high-energy electron scattering at the future Electron-Ion Collider (EIC), focusing on nuclear fragmentation processes induced by photon energies in the range of giant resonances ($\sim 10$–$20$~MeV)~\cite{BertulaniKucuk2025}. In contrast, the present work discusses the potential for investigating neutron skins using high-energy photons in the hundreds of MeV to GeV range. This approach may offer a complementary perspective on neutron skin properties and their connection to the nuclear equation of state.
Our analysis is motivated by  the  studies performed in Refs.~\cite{Sengul:2015ira,Xu:2024dja}, focused on the dilepton production by $\gamma\gamma$ interactions in UPCs, that have pointed out the potentiality of UPCs to probing the neutron skin. As we will demonstrated below, the photoproduction of vector mesons in UPCs is also sensitive to the presence (or not) of a neutron skin. It is worthwhile mentioning that nuclear imaging probed with  vector meson photoproduction is a central part of the physics programs of the EIC \cite{Asche19} and has been studied theoretically in UPCs at the LHC and in deep inelastic (DIS) electron scattering at the EIC \cite{Maanfa22,Maanfa23}.

This paper is organized as follows. In Section~II, we introduce a quantitative measure of the localization of photonuclear interactions, referred to as the photonuclear tomography function. Moreover,  a brief overview of the vector meson content in photons is presented. In Section~III, we present our numerical results, focusing primarily on the production of $J/\psi$ and $\rho^0$  mesons in UPCs. Finally, Section~IV summarizes the general conclusions of this work.

\begin{figure}[t]
\begin{center}
\includegraphics[scale=0.46]{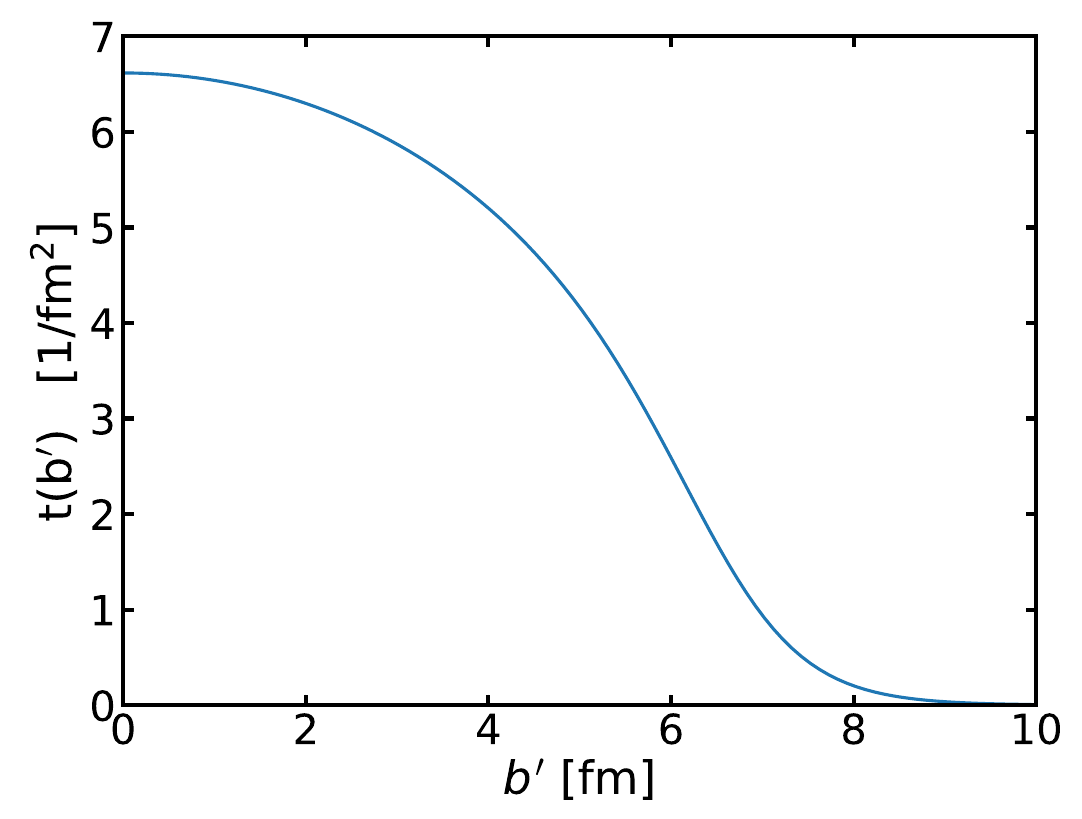}
\caption{
The thickness function defined in Eq.~\eqref{thick} for Pb nuclei as a function of the transverse distance $b'$ from the nuclear center.  
 \label{EICws01}}
\end{center}
\end{figure}

 \section{Photonuclear tomography in ultraperipheral collisions (UPC)}
 \subsection{UPC in coordinate space \label{susec:UPCcs}} 
When two relativistic ions pass each other at distances larger than the sum of their radii, they interact primarily via their electromagnetic fields - a process known as ultraperipheral heavy-ion collisions (UPCs). For relativistic ions with $Z \gg 1$, the electric and magnetic fields are predominantly perpendicular to the direction of motion of the ions. This field configuration acts as a source of quasi-real (almost real) photons, resulting in an intense photon flux~\cite{ISI:A1988P121800001,annurev.nucl.55.090704.151526}. In this work, we consider one of the ions (ion $A$) as the source of the photon flux and study the interaction of these photons with the opposing nucleus (ion $B$).

High-energy photons from the equivalent photon spectrum of ion $A$ interact with individual nucleons within the nucleus $B$ inside a transverse area element $db^{\prime 2}$. The number of nucleons available for interaction in ion $B$ is determined by the nuclear ground-state density, $\rho_B({\bf r'})$, through the so-called nuclear thickness function:
\begin{equation}
t_{B}({\bf b}') = \int dz' \rho_B({\bf r'}), \label{thick}
\end{equation}
where ${\bf r'} = ({\bf b}', z')$, with $z'$ denoting the coordinate along the beam (collision) axis and ${\bf b}'$ the transverse position within the nucleus. Figure \ref{EICws01} displays the thickness function calculated for Pb with the density parameters described below.

\begin{figure}[b]
\begin{center}
\includegraphics[scale=0.42]{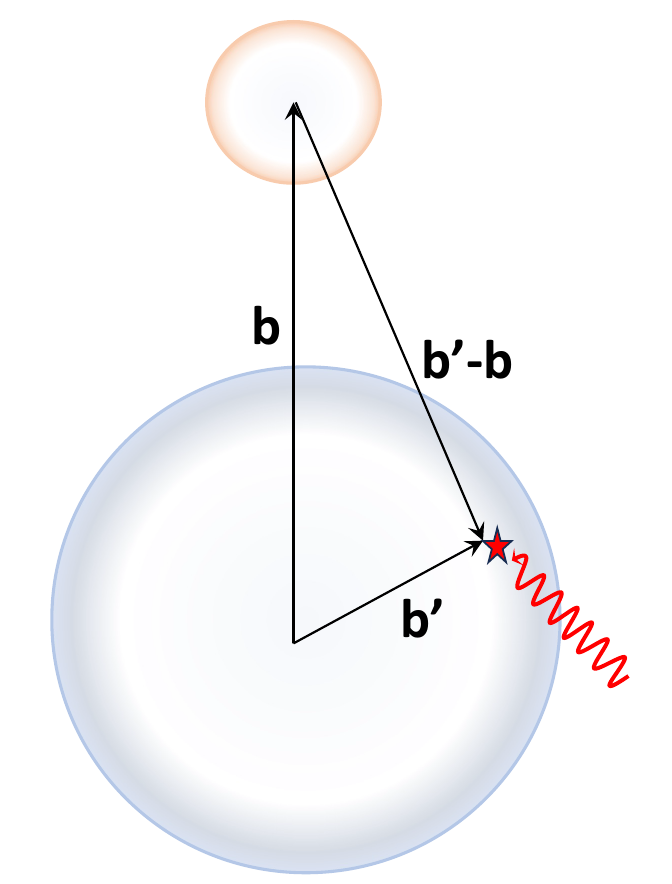}
\caption{Nuclear collision as viewed perpendicularly to the beam direction. The impact parameter ${\bf b}$ is shown as well as the transverse distance ${\bf b'}$ from the nuclear center where the quasi-real photon strucks. $|{\bf b} - {\bf b'}|$ is the distance from the center of the ``projectile".\label{colli}}
\end{center}
\end{figure}

The photon flux per unit area and per unit energy incident at a point ${\bf b}'$, originating from the ion $A$ with Lorentz boost factor $\gamma \gg 1$, is well established~\cite{ISI:A1988P121800001,annurev.nucl.55.090704.151526}. By multiplying this flux by the nuclear thickness function - representing the probability of finding a nucleon at position ${\bf b}'$ - we define the photonuclear interaction distribution function:
\begin{equation}
\frac{dN_\gamma (\kappa, {\bf b}, {\bf b}')}{d\kappa\, d^2b\, d^2b'} 
= \frac{Z_A^2 \alpha \kappa}{\pi^2 \gamma^2} 
K_1^2\left( \frac{\kappa |{\bf b} - {\bf b}'|}{\gamma} \right) 
t_{B}({\bf b}') S({\bf b}), \label{PF}
\end{equation}
where $\kappa$ is the photon energy, and ${\bf b}$ is the impact parameter between the centers of the colliding ions. This expression assumes spherical charge distributions, consistent with Birkhoff's theorem. The factor $S({\bf b})$ represents the ion-ion survival probability and is given by
\begin{equation}
S({\bf b}) = \exp\left[ - \sigma_{NN} \int dz \int d^3r' 
\rho_{A} ({\bf r}) \rho_{B} ({\bf r} - {\bf r}') \right], \label{sb}
\end{equation}
where ${\bf r} = ({\bf b}, z)$, $\rho_{i}({\bf r})$ denotes the ground-state density of ion $i = A, B$, and $\sigma_{NN}$ is the nucleon-nucleon total cross section, assumed to be isospin independent. Figure \ref{colli} displays a schematic view of the nuclear collision perpendicular to the beam direction. The impact parameter ${\bf b}$ is shown as well as the transverse distance ${\bf b'}$ from the nuclear center where the quasi-real photon strucks. $|{\bf b} - {\bf b'}|$ is the distance from the center of the ``projectile".

Since the photons are quasi-real, Eq.~\eqref{PF} includes only the contribution from transversely polarized photons, which dominate in this regime.
The Lorentz factor $\gamma$ in Eq.~\eqref{PF} is calculated in the rest frame of one of the colliding ions. For lead-lead (Pb+Pb) collisions at the LHC with center-of-mass energy $\sqrt{s_{NN}} = 5.5$~TeV, the laboratory beam energy is $E_{\mathrm{lab}} = 2.76$~TeV per nucleon, corresponding to a Lorentz factor $\gamma_{\mathrm{lab}} = 2,941$. In the rest frame of one of the ions (ion $B$), the Lorentz factor for the other ion becomes $\gamma = 2 \gamma_{\mathrm{lab}}^2 - 1 \approx 1.73 \times 10^7$.

The minimum value of $|{\bf b} - {\bf b}'|$ in Eq.~\eqref{PF} is approximately the nuclear radius of lead, $R_{\mathrm{Pb}} \simeq 6.5$~fm. The function $x K_1(x)$ approaches unity for $x \lesssim 1$ and decays exponentially for $x \gtrsim 1$. Even for the heaviest known vector meson, with $m_V \sim 6$~ GeV, the argument for the Bessel function remains small for the most relevant range of impact parameters, $15 \lesssim b \lesssim 10^5$~ fm. Within this range, the right-hand side of Eq.~\eqref{PF} approximately scales as $1/|{\bf b} - {\bf b}'|^2$ for $b \gtrsim 2 R_{\mathrm{Pb}}$.

\begin{figure}[t]
\begin{center}
\includegraphics[scale=0.5]{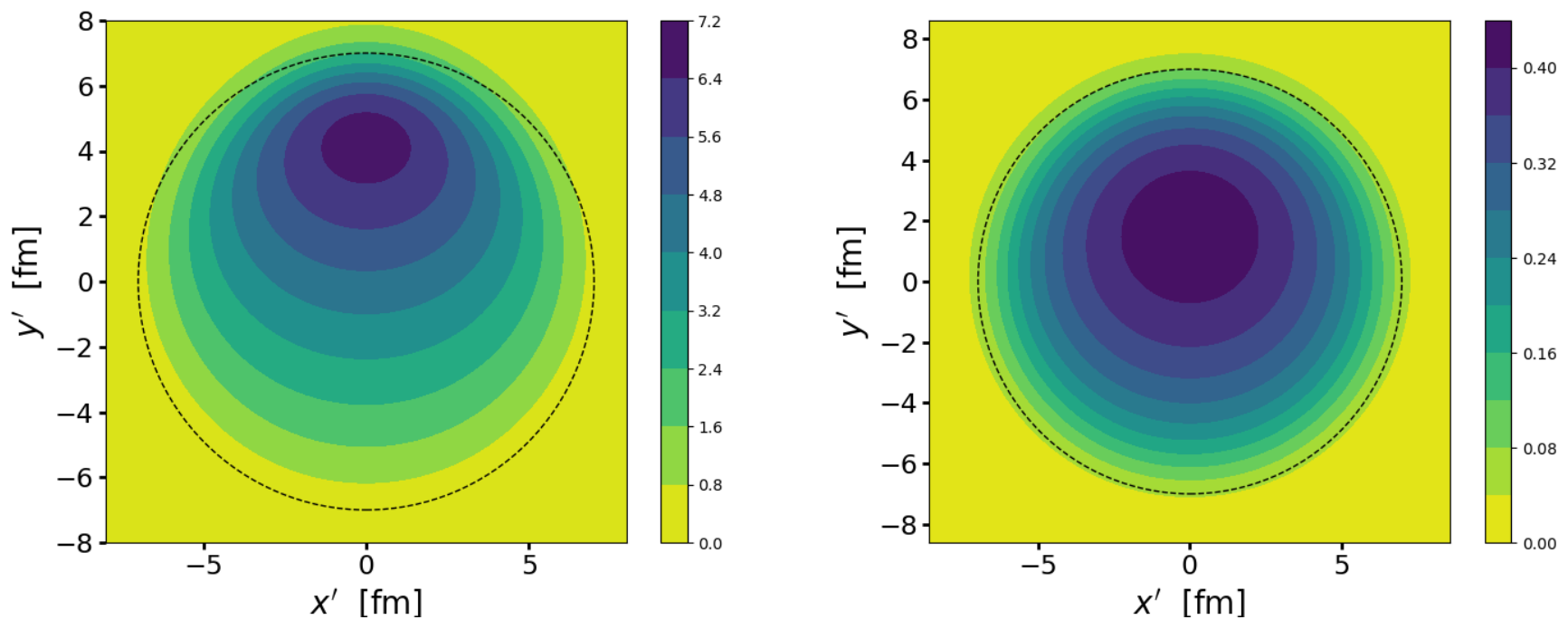}
\caption{The function ${dN_\gamma (\kappa,{\bf b},{\bf b}') / d\kappa\, d^2b\, d^2b'}$ from Eq.~\eqref{PF} evaluated at fixed photon energy $\kappa = 6$~GeV for two different impact parameters: $b = 15$~fm (left panel) and $b = 40$~fm (right panel). The contour plots display iso-surfaces of equal function values in the transverse plane $(x', y')$ relative to the nuclear center at $(0,0)$. Darker regions correspond to higher values of the function. The dashed circle has a radius of 7 fm, enclosing the bulk of the nuclear mass distribution. \label{EICwsN}}
\end{center}
\end{figure}

To better illustrate the effective spatial distribution of photons interacting with the nucleus, Fig.~\ref{EICwsN} shows contour plots of the function ${dN_\gamma (\kappa,{\bf b},{\bf b}') / d\kappa\, d^2b\, d^2b'}$ for a fixed photon energy $\kappa = 6$~GeV at two impact parameters, $b = 15$~fm $(\simeq 2 R_{\mathrm{Pb}})$ and $b = 40$~fm. The plots display contours of equal flux values across the transverse plane within the Pb nucleus, with $(x', y')$ measured relative to the nuclear center, $b' =\sqrt{x'^2 + y'^2}$. The dashed circle has a radius of 7 fm, enclosing the bulk of the nuclear mass distribution.

The results clearly demonstrate that for grazing collisions ($b \simeq 2 R_{\mathrm{Pb}}$), the flux of high-energy photons is concentrated near the side of the nucleus closest to the passing ion. As the impact parameter increases, photons predominantly interact near the nuclear center. Consequently, different regions of the nucleus are probed depending on the impact parameter. For large impact parameters ($b \gtrsim 100$~fm), the central region of the nucleus is most strongly illuminated by the photon flux, with the interaction strength falling to half its maximum value at approximately 2/3 of the nuclear radius. This behavior reflects the dependence of the thickness function on the transverse distance from the nuclear center.

\subsection{Photonuclear tomography function}
If we integrate Eq. \eqref{PF} over all impact parameters, the resulting function depends solely on the magnitude of ${\bf b'}$, assuming spherically symmetric nuclear densities. By multiplying this result with cross section for the photon - nucleus $B$ interaction, $\sigma_{\gamma B}(\kappa)$, for a given photon energy $\kappa$, we define the Photonuclear Tomography Function (PTF). The PTF represents the probability that a photon will interact with the nucleus at a transverse distance $b'$ from its center and produce a meson. This is given by
\begin{equation} {\cal P}_T (\kappa, b') = \frac{Z_A^2 \alpha \kappa^2}{\pi^2 \gamma^2} \, \sigma_{\gamma B \rightarrow VB}(\kappa) \, t_B(b') \int d^2b \, K_1^2\left( \frac{\kappa |{\bf b} - {\bf b'}|}{\gamma} \right) S(b).
\label{phiN} \end{equation}
This expression can be generalized straightforwardly to deformed nuclei, such as in uranium-lead (U+Pb) or uranium-uranium (U+U) collisions. It is important to distinguish the PTF from the probability of a photonuclear reaction occurring in an ion-ion collision at a specific impact parameter $b$. The PTF instead reflects the probability that a photon interacts with the nucleus $B$ at a given distance $b'$ from its center, after integrating over all impact parameters. This feature makes the PTF particularly relevant for heavy-ion collision experiments, where UPC events are typically not resolved by a specific value of $b$.

The total photoproduction cross section for a vector meson in UPCs is then obtained by integrating the PTF over both the photon energy $\kappa$ and the transverse position $b'$: \begin{equation} \sigma(AB \rightarrow A \otimes V \otimes B) = 2\pi \int \frac{d\kappa}{\kappa} \int db' \, b' \, {\cal P}_T(\kappa, b')\,\,, \end{equation}
where $\otimes$ represents the presence of a rapidity gap in the final state.

The formalism developed here applies broadly to photonuclear processes induced in UPCs. At low photon energies - such as those associated with the excitation of giant resonances - the photon wavelength becomes comparable to or larger than the nuclear radius, making the localization of the photon within the nucleus less well-defined. Nevertheless, the same formalism remains valid for other meson production channels, including pion photoproduction processes like $\gamma N \rightarrow N \pi^0$ and $\gamma N \rightarrow N \pi^+ \pi^-$, which are predominantly driven by $\Delta$ resonance excitation and decay. 
For vector meson production, the situation is different: the photon wavelength is typically  smaller than the nucleon size. For instance, in the case of $J/\psi$ production, the photon wavelength satisfies $\lambda = 2\pi / M_V \lesssim 0.3$ fm. Under these conditions, the number of nucleons available for interaction with the photon within a transverse area $d^2 b'$ is effectively encoded in the nuclear thickness function $t({\bf b'})$.

\subsection{The photon as a vector meson}
In the Vector Meson Dominance (VMD) model, photonuclear interactions are described in terms of the photon fluctuating into vector meson states~\cite{1960AnPhy..11....1S}. According to the uncertainty principle, the lifetime of such fluctuations is given by $2\kappa/(Q^2 + M_V^2)$, where $\kappa$ is the photon energy, $Q$ is the momentum transfer, and $M_V$ is the vector meson mass. This fluctuation lifetime increases with beam energy. For instance, a photon beam with $\kappa = 50$~GeV fluctuates into a $\rho$ meson state over a distance of approximately 4~fm, comparable to the typical nuclear dimensions. Consequently, the interaction of high-energy photons with matter resembles hadronic interactions.

The photon wave function can be expressed as~\cite{PAUL1985203,SCHULER1993539}
\begin{equation}
\big| \gamma \big>
= \left[ c_0 \big| \gamma_0 \big> 
+ \sum_V \frac{e}{f_V} \big| V \big> 
+ \sum_q c_q \big| q\bar{q} \big> \right], 
\label{gamexp}
\end{equation}
where $\big| \gamma_0 \big>$ denotes the point-like photon state, $\big| V \big>$ represents the vector meson state, and $\big| q\bar{q} \big>$ is a generic quark-antiquark pair state. The coefficients $c_0$, $e/f_V$, and $c_q$ describe the respective couplings of these components. At high photon energies, fluctuations such as $\gamma \rightarrow J/\psi$ can be treated perturbatively~\cite{SCHULER1993539}. The probability for the transition $\gamma \rightarrow V$ is proportional to $(e/f_V)^2 = 4\pi \alpha / f_V^2$. Therefore, in the VMD framework, the equivalent photon flux in UPCs effectively corresponds to a beam of vector mesons with energy-dependent composition across the virtual photon spectrum.

The cross section for the production of a specific vector meson is determined by two main factors:  
(a) the number of equivalent photons at a given photon energy $\kappa$, and  
(b) the coupling coefficients in the expansion~\eqref{gamexp} at the same energy $\kappa$.
The coupling strength $f_V$ in Eq.~\eqref{gamexp} reflects the photon-vector meson coupling and depends on the sum of the electric charges of the quarks within the meson:
$
f_V \propto 1/\sum_{q_V} e_{q_V}.
$
It can be extracted from the electronic decay width of the corresponding vector meson via
\begin{equation}
\Gamma_{V \rightarrow e^+e^-} 
= \frac{8\pi \alpha^2}{3} 
\frac{f_V^2}{m_V} 
\left( \sum_{q_V} e_{q_V} \right)^2. 
\label{Gee}
\end{equation}
For example, the factor $2(\sum_{q_V} e_{q_V})^2$ takes the values $(1, \tfrac{1}{9}, \tfrac{2}{9}, \tfrac{8}{9}, \tfrac{2}{9})$ for $V = (\rho^0,  \omega, \phi, J/\psi, \Upsilon)$, respectively, and the corresponding coupling constants $f_V$ (in GeV) are $(0.13, \ 0.11,  \ 0.12,\ 0.20,\ 0.38)$~\cite{Yin2021-qj}.

When the photon interacts with a nucleon via its $\big| V \big>$ component, elastic scattering on the gluon field of the nucleon or nucleus can occur, producing the vector meson $V$ as an observable final state. This process has been proposed as a probe of the gluon distribution in nuclei through vector meson production in UPCs~\cite{PhysRevC.65.054905}. In contrast, the point-like photon component $\big| \gamma_0 \big>$ can interact via QED processes with target quarks, leading to hadronic jet production. However, this contribution represents less than 1\% of the total $\gamma N \rightarrow$ hadrons cross section and is typically neglected~\cite{PAUL1985203,SCHULER1993539}.

The mean free path of the virtual vector meson within the nucleus depends on the meson species. For light mesons like the $\rho^0$, the mean free path is approximately 1~fm ($\sigma_{\rho^0 N} = 40$~mb), whereas for heavier mesons like the $J/\psi$, it can extend to several femtometers ($\sigma_{J/\psi N} = 5$~mb). Theoretically, the effective cross section must also account for shadowing effects, which modify the number of nucleons effectively participating in the interaction. These effects are described by an effective nucleon number $A_{\mathrm{eff}}$, which may differ from the actual nucleon number $A$. However, studies of high-energy photon-nucleus interactions at the quark-gluon level suggest that $A_{\mathrm{eff}} \approx A$ remains a reasonable approximation~\cite{PhysRevC.71.054902}.

\begin{figure}[t]
\begin{center}
\includegraphics[scale=0.41]{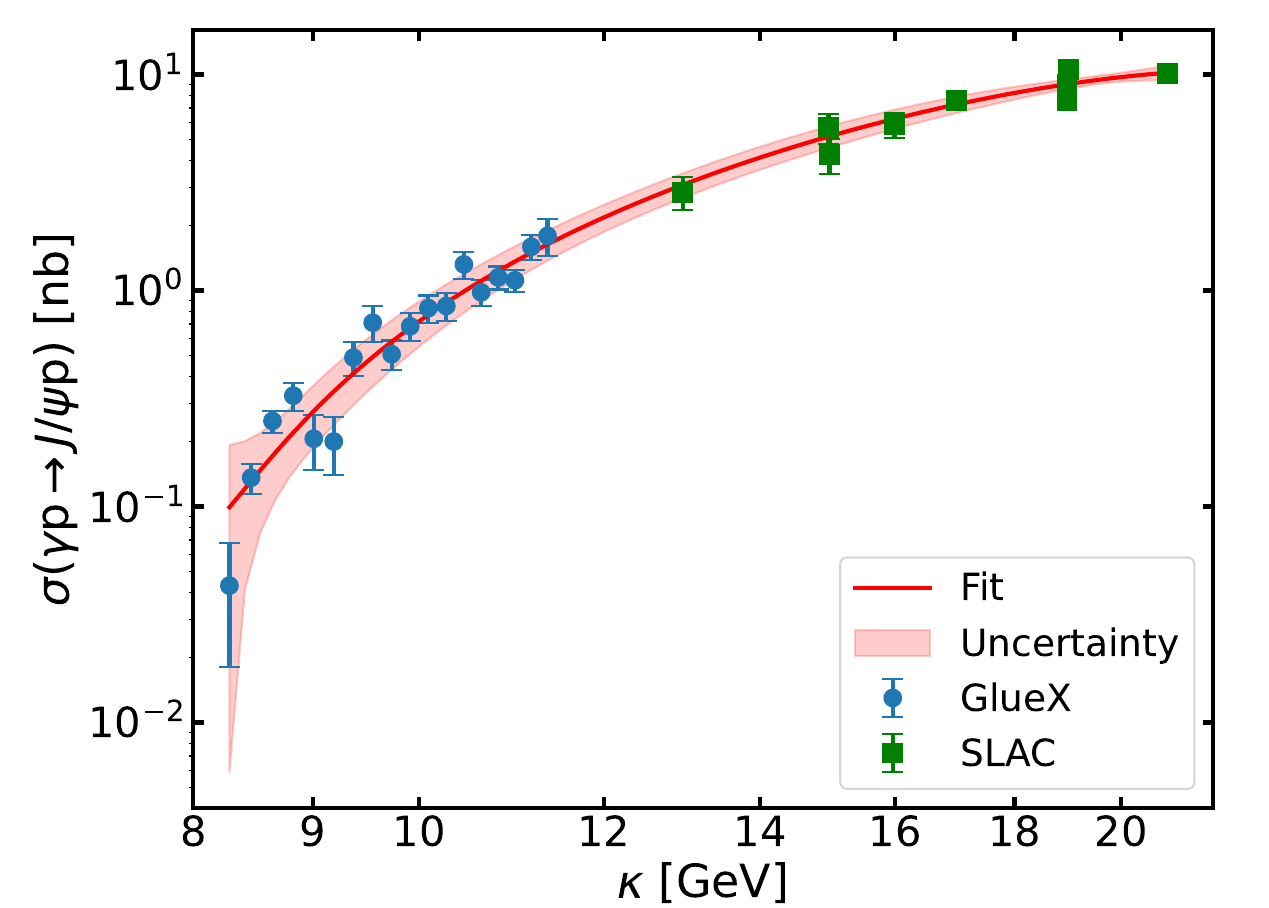}
\caption{
 Experimental data for $\sigma_{\gamma p \rightarrow J/\psi p}$, showing $J/\psi$ production from SLAC~\cite{PhysRevLett.35.483} and JLab~\cite{PhysRevC.108.025201}. A polynomial fit to the data is performed using $\sigma\ \mathrm{(nb)} = a\kappa + b\kappa^2 + c\kappa^3$, where $\kappa$ is the photon energy in GeV. The fit parameters are $(a, b, c) = (0.298 \pm 0.027, -3.429 \pm 0.338, 11.802 \pm 1.336)$. The uncertainty band is derived from the covariance matrix of the fit parameters. \label{EICws02}}
\end{center}
\end{figure}

\section{Numerical Results}
\label{sec:results}
In order to estimate the photonuclear tomography function, we must assume a model for the cross section $\sigma_{\gamma B \rightarrow VB}$. Such a quantity is commonly related to experimental data for the vector meson photoproduction in photon - proton interactions  through the expression
\begin{equation} 
\sigma_{\gamma B \rightarrow VB} = \left. \frac{d \sigma_{\gamma p \rightarrow Vp}}{dt} \right|_{t=0} \int_{t_{\text{min}}}^\infty dt \,  F_B(t), \label{aft} 
\end{equation} 
where $d \sigma_{\gamma p \rightarrow Vp} / dt \big|{t=0}$ represents the forward scattering amplitude, and $F_B(t)$ is the nuclear form factor of nucleus $B$. The forward scattering amplitude contains the essential dynamical information, while the form factor accounts for the elastic scattering dependence on momentum transfer. The minimum longitudinal momentum transfer required for coherent particle production is given by $t_{\text{min}} = -p^2_{z_{\text{min}}} = -M_V^4 / (4 \kappa \gamma_L)^2$.

However, when studying nuclear tomography - such as in $J/\psi$ production via UPCs - it is not necessary to rely on Eq. \eqref{aft} or to construct a theoretical model for $\sigma_{\gamma p \rightarrow Vp}$. A wealth of experimental data is now available from SLAC \cite{PhysRevLett.35.483}, Jefferson Lab \cite{PhysRevC.108.025201}, HERA \cite{Alexas2013, Chekanov2002}, and recent LHC UPC measurements \cite{Acharya2019}. By directly fitting these data for $\sigma_{\gamma p \rightarrow Vp}$, we can reliably estimate the relevant cross sections and probabilities. Additionally, the transparency function introduced in Eq. \eqref{thick} already accounts for the effective number of participating nucleons.

Figure \ref{EICws01}  illustrates the nuclear thickness function, as defined in Eq. \eqref{thick}, for lead nuclei as a function of the transverse distance $b'$ from the nuclear center. Although the integral of this function corresponds to the total nucleon number, Eq. \eqref{aft} oversimplifies the photon-nucleus interaction by assuming a uniform photon interaction across the nucleus. As we demonstrate below, the Photonuclear Tomography Function (PTF) defined in Eq. \eqref{phiN} provides a more accurate description of the spatial localization of the photon-nucleus interaction.

Using Eq. \eqref{phiN} for vector mesons $V = (\rho^0,  \omega, \phi, J/\psi, \Upsilon)$, the integration over the impact parameter can be performed analytically with a sharp cutoff approximation for the minimum impact parameter. The PTF at the center of the nucleus ($b' = 0$) is given by
\begin{equation} {\cal P}_T (\kappa, 0) = \frac{2 Z_A^2 \alpha}{\pi} \sigma_{\gamma p \rightarrow Vp} (\kappa)  t_B(0) \ln \left( \frac{\delta \gamma}{\kappa D} \right), \label{phiNb} \end{equation} where $\delta \approx 0.68108$ (related to Euler's constant), and $D \simeq 2.4 A^{1/3}$ fm represents the average distance between colliding nuclei at the minimum impact parameter in UPCs. The logarithmic factor $\ln \left( \delta \gamma / \kappa D \right)$ is identical for any meson produced by a photon of energy $\kappa$, thus highlighting the validity of the equivalent photon approximation in determining the expansion coefficients in Eq. \eqref{gamexp}. This allows for extracting ratios of production cross sections, $\sigma_{\gamma p \rightarrow V_2 p} / \sigma_{\gamma p \rightarrow V_1 p}$, for different mesons at the same photon energy and collision conditions. If perturbative photon exchange is not applicable, however, direct comparisons between vector meson production in UPCs and real-photon data require additional modeling beyond the equivalent photon approach.

We now turn to $J/\psi$ production near threshold, which is considered a promising tool to investigate the $J/\psi$ nucleon scattering length - believed to be related to two-gluon matrix elements and the trace anomaly contribution to the nucleon mass \cite{PhysRevLett.74.1071}. Figure \ref{EICws02}  presents a polynomial fit of the SLAC and GlueX data using
$\sigma \ {\rm (nb)} =  a\kappa + b\kappa^2 + c\kappa^3$ with $\kappa$ in GeV. The fit yields parameters $(a, b, c) = (0.298 \pm 0.027, -3.429 \pm 0.338, 11.802 \pm 1.336)$, and the uncertainty band is derived from the covariance matrix of the fit. For $\kappa > 25$ GeV, we adopt a power-law parametrization, $\sigma_{\gamma p \rightarrow Vp } = \sigma_0 W_{\gamma p}^\delta$ , inspired by previous analyses of ALICE data \cite{Alexas2013}. Here, $W_{\gamma p}$ is the invariant mass of the photon-proton system, related to the rapidity $y$ of the $J/\psi$ produced by $W_{\gamma p}^2 = 2 E_p M_{J/\psi} e^{-y}$, with $E_p = 2.76$ TeV. The photon energy in the target frame is given by $\kappa = (M_{J/\psi} / 2) e^{y}$. Using data from HERA and ALICE \cite{Alexas2013, Chekanov2002} in the range $25 < W_{\gamma p} < 305$ GeV, we obtain $\sigma_0 = 3.75 \pm 0.05$ nb and $\delta = 0.677 \pm 0.08$. These fits are used for our predictions of the PTF and total $J/\psi$ production cross sections in UPCs.

\begin{figure}[t] \begin{center} 
\includegraphics[scale=0.35]{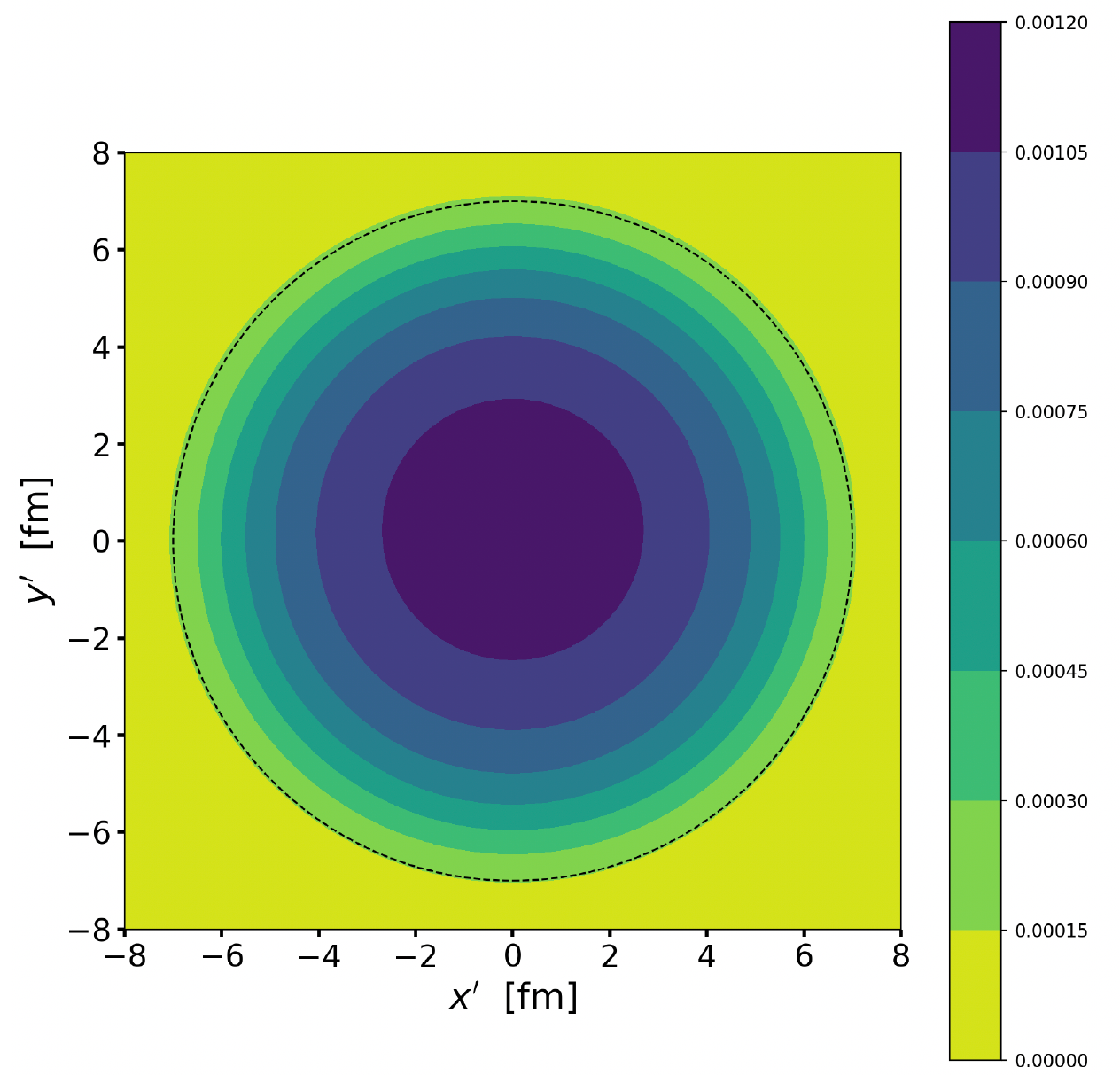}  \ \ \ \ 
\includegraphics[scale=0.50]{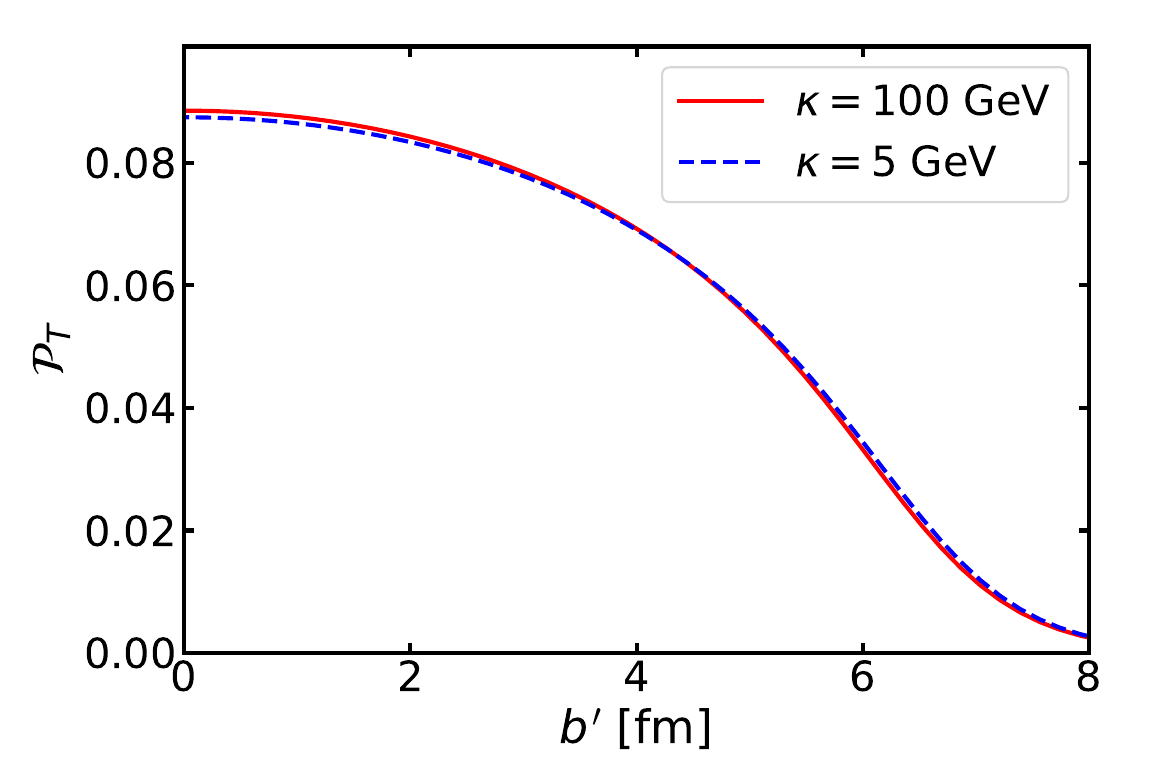} 
\caption{Left: Contour plot of the photonuclear Tomography Function (PTF) from Eq. \eqref{phiN} as a function of the transverse coordinate ${\bf b'} = (x',y')$ from the nuclear center, for $J/\psi$ production in Pb + Pb collisions at CERN energies at photon energies  $\kappa = 10$ GeV. Right: The same function  as a function of $b^\prime$ for $\kappa=5$ GeV (dashed line) and 100 GeV (solid line). For clarity, the two PTFs have been normalized to be visible in the same scale as for $\kappa = 100$ GeV. 
\label{gJPsi2}} \end{center} \end{figure}

Figure \ref{gJPsi2} displays on the left panel the contour plot of the Photonuclear Tomography Function (PTF) from Eq. \eqref{phiN} as a function of the transverse coordinate ${\bf b'} = (x',y')$ from the nuclear center, for $J/\psi$ production in Pb + Pb collisions at CERN energies at photon energies $\kappa = 10$ GeV. For higher photon energies, such as $\kappa \sim 100$ GeV, the cross section $\sigma_{\gamma p \rightarrow J/\psi p}$ increases by two orders of magnitude, yielding ${\cal P}_T (\kappa, 0) \sim 0.1$.  This confirms the applicability of the one-photon exchange approximation even at high photon energies, with higher-order effects not due to photon-hadron interactions but because of hadron-hadron rescattering after the initial photonuclear interaction.  In the figure on the right panel we plot the same function as a function of $b^\prime$ for $\kappa=5$ GeV (dashed line) and 100 GeV (solid line).  For clarity, the two PTFs have been normalized to be visible on the same scale as for $\kappa = 100$ GeV.  The PTFs for the different photon energies are very similar, with the highest photonuclear interaction probabilities occurring at the nuclear center, where the production probability reaches approximately $10^{-3}$ ($\kappa=5$ GeV) and $0.1$ ($\kappa=100$ GeV). These values decrease by half at $b' \approx 5$ fm, reflecting the reduced nucleon density near the nuclear surface.  
The reason for the very weak dependence of the width of PTF with photon energy $\kappa$ is the same as that discussed in Section \ref{susec:UPCcs}. Only for very large impact parameters $b$, of the order of 10,000 fm or more, the argument of the modified Bessel function in the integrand of Eq. \eqref{phiN} will display a stronger dependence on the internal nuclear distance $b'$. At these large values most of the integral of the ion-ion impact parameter has been accounted for.

\begin{figure}[t] \begin{center} 
\includegraphics[scale=0.4]{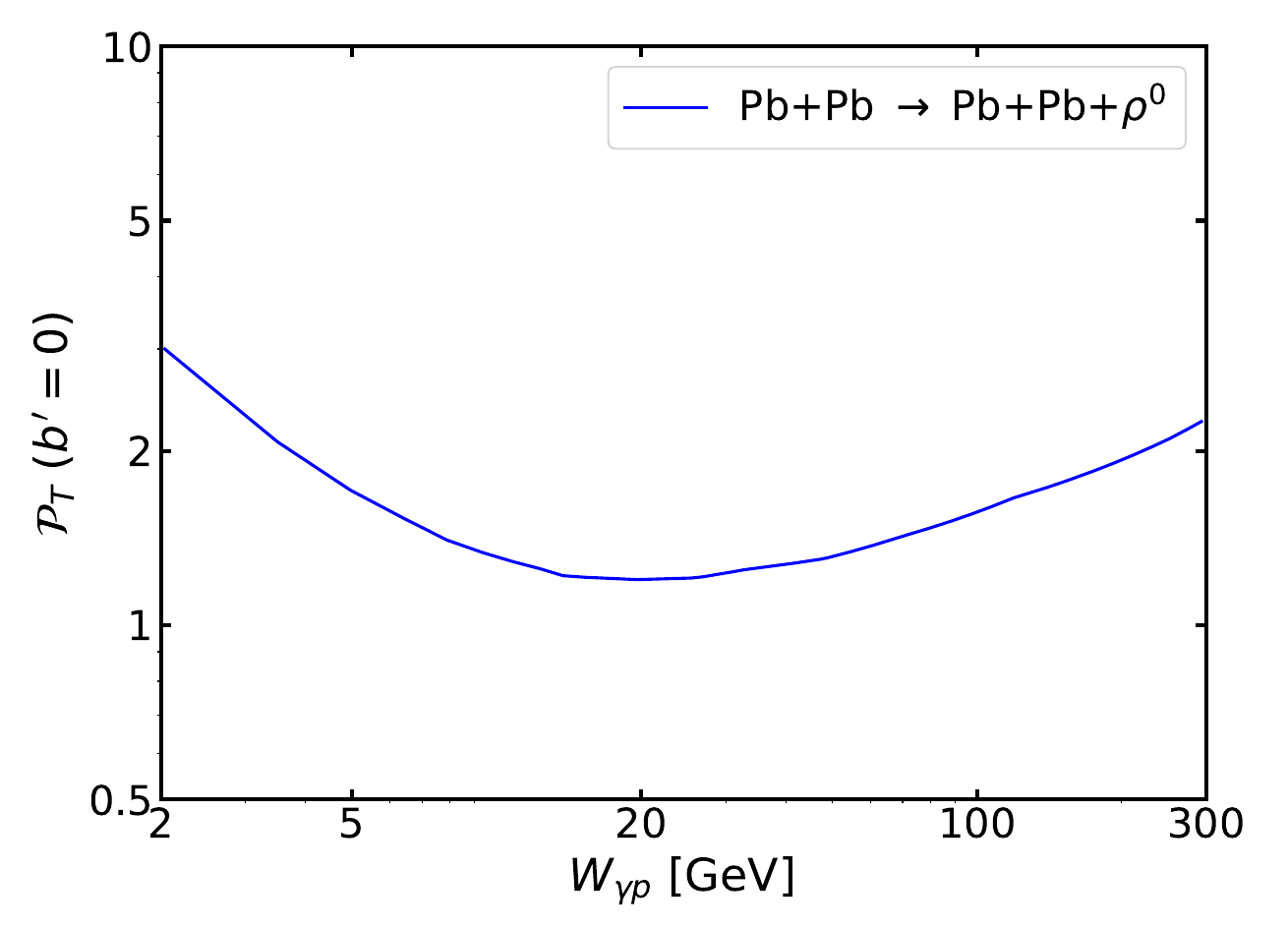} 
\caption{The  photonuclear Tomography Function (PTF) for the production of a $\rho^0$ meson at the center of one of the Pb nuclei ($b'=0$) as a function of the invariant mass $W_{\gamma p}$.
\label{rho}} \end{center} \end{figure}

As previously discussed, higher-order hadronic scattering corrections are expected in certain cases - particularly for $\rho^0$ production because of its large nucleon scattering cross section, $\sigma_{\rho^0 N} \sim 40$ mb. In such scenarios, even the primary photon-exchange process may require a perturbative treatment beyond the leading order. Experimental data for $\rho^0$ photoproduction on proton targets exhibit a peak near $W_{\gamma p} = 2$ GeV \cite{PhysRevLett.89.272302}. These data are well described by a phenomenological parameterization of the form
$\sigma_{\gamma p \rightarrow p + \rho} =\alpha_1 W_{\gamma p}^{\delta_1} + \alpha_2 W_{\gamma p}^{\delta_2}$, with fit parameters $\delta_1 = -0.81 \pm 0.06$ and $\delta_2 = 0.36 \pm 0.06$ keeping the other parameters fixed, i.e., $\alpha_1= 33.65$ $\mu$b and $\alpha_2 = 1.986$ $\mu$b \cite{Sirunyan-rho-prod-2019}. Using the maximum value $\sigma_{\gamma p \rightarrow p \rho^0} \approx 20\ \mu$b at $W_{\gamma p} = 2$ GeV, we estimate that the PTF at the center of the nucleus for $\rho^0$ production in Pb + Pb collisions at LHC energies reaches approximately 2.9. This is better seen in Fig. \ref{rho} where we plot the PTF for the production of a $\rho^0$ meson at the center of one of the Pb nuclei ($b'=0$) as a function of the invariant mass $W_{\gamma p}$. The high values of the PTF indicate that the one-photon exchange approximation is insufficient, and higher-order photon exchanges must be considered. Consequently, the photoproduction cross sections for $\rho^0$ mesons extracted from UPC data using the equivalent photon method may require reanalysis to properly account for these effects.

As emphasized in the introduction, a major motivation for studying photonuclear reactions is their potential to probe the neutron skin of heavy nuclei, which is relevant for constraining the equation of state (EoS) of neutron-rich matter and neutron stars. At GeV energies, vector-meson production in UPCs presents an abundant reaction channel that could, in principle, be sensitive to neutron skin effects. However, as already suggested by Fig. \ref{gJPsi2}, the photonuclear interaction probability is suppressed at the nuclear surface, where the neutron skin resides. This suppression results from the reduced nucleon density at large transverse distances.

Despite this, absorption effects at small impact parameters can still influence the PTF at the nuclear center. Since small impact parameters enhance the energy of the quasi-real photons, the central PTF remains sensitive to the neutron skin, although indirectly. If we assume that the photo-production cross section $\sigma_{\gamma p \rightarrow Vp}(\kappa)$ is identical for protons and neutrons, then the isospin dependence of the absorption factor $S(b)$ in Eq. \eqref{phiN} can be expressed as a product of proton and neutron contributions: \begin{equation} S(b) = S_p(b) S_n(b), \end{equation} where 
\begin{equation} S_p({\bf b}) = \exp\left[ - \sigma_{pp} \int dz \int d^3r' \rho_{A}^{(p)}({\bf r}) \rho_{B}^{(p)}({\bf r - r'})
-\sigma_{pn} \int dz \int d^3r' \rho_{A}^{(p)}({\bf r}) \rho_{B}^{(n)}({\bf r - r'}) \right]. \label{sb2} 
\end{equation} 
Here, $\rho_{A}^{(p)}$ and $\rho_{A}^{(n)}$ denote the ground-state proton and neutron densities, respectively. A similar expression applies for $S_n(b)$ by interchanging protons and neutrons ($p \leftrightarrow n$).

The inelastic proton-proton interaction cross section at LHC energies has been measured by the ATLAS collaboration at $\sqrt{s} = 7$ TeV, yielding $\sigma_{pp} = 60.3 \pm 2.1$ mb \cite{10.1038/ncomms1472}. We adopt this value as representative for nucleon-nucleon interactions in Pb + Pb collisions at $\sqrt{s_{NN}} = 5.5$ TeV, although extrapolations to this energy carry significant uncertainties, as discussed in Ref. \cite{10.1038/ncomms1472}. Since no data exists for neutron-proton ($np$) and neutron-neutron ($nn$) cross sections at these energies, we assume for simplicity that $\sigma_{nn} = \sigma_{pn} = \sigma_{pp}$. This assumption reduces the model dependence of the neutron skin analysis through absorption effects.

The neutron skin primarily influences the nuclear absorption at small impact parameters. For $b' = 0$, the logarithmic dependence in Eq. \eqref{phiNb} implies that variations in the nuclear radius parameter $D$ (e.g., by 5\%) result in changes to ${\cal P}_T (\kappa, 0)$ of less than 1\%. However, at finite $b'$, the sensitivity to the neutron skin becomes more significant, as the quasi-real photons probe the outer regions of the nucleus, where the neutron excess is located - consistent with the behavior shown in Fig. \ref{EICwsN} (left panel). Consequently, the total cross section for vector-meson production retains some sensitivity to neutron skin effects.

To explore this sensitivity quantitatively, we model the charge distribution in Pb as a Fermi function with radius and diffuseness parameters $R_0 = 6.62$ fm and $a = 0.546$ fm, respectively. Assuming the same parameters for the neutron distribution yields a cross section for $J/\psi$ production in Pb + Pb collisions at LHC energies (over $4 < W_{\gamma p} < 305$ GeV) of $48 \pm 1.5$ mb, where the uncertainty reflects the error in $\sigma_{\gamma p \rightarrow J/\psi p}(\kappa)$. If, instead, we assume that the neutron distribution has a larger rms radius with a neutron skin thickness $R_{np} = \sqrt{\langle r_n^2 \rangle - \langle r_p^2 \rangle} = 0.2$ fm, the resulting cross section decreases slightly to $46 \pm 1.4$ mb - a difference of approximately 4\%.
This modest change highlights the experimental challenge of using vector-meson production in UPCs as a precise probe of neutron skin properties. Additionally, uncertainties in the elementary nucleon-nucleon cross sections at these energies further complicate this approach. Alternative channels, such as $\pi^+ \pi^-$ pair production in UPCs, also suffer from significant pion-nucleon rescattering effects, limiting their effectiveness as neutron skin probes.

\section{Conclusions}
\label{sec:conclusion}

In this work, we have developed a theoretical framework for photonuclear tomography as a means to investigate the internal structure of nuclei using ultraperipheral heavy-ion collisions (UPCs). By exploiting the intense flux of virtual photons generated in UPCs, this method enables coherent photonuclear vector meson production that are sensitive to the spatial distribution of nuclear transition densities.

Through the use of the equivalent photon approximation, we have calculated photon fluxes and photonuclear cross sections for representative nuclear systems. The analysis of the spatial dependence of particle production within the nucleus allows for a clearer physical insight of the process. The dependence of the photonuclear particle production on the nuclear geometry provides crucial information about the spatial characteristics of nuclear excitations. This can also be of relevance for deformed nuclei  such as in uranium + uranium collisions.

Photonuclear tomography offers several advantages as a probe of nuclear structure, particularly for heavy nuclei where coherent interactions enhance the sensitivity to meson masses. With proper modifications, this technique may also be applied to electron-ion colliders, providing valuable insights into regions of the nuclear chart that are difficult to access with traditional methods.
Future work will focus on extending the application of this framework to other nuclear systems and exploring higher multipolarities. Experimental validation of the predicted observables may further establish photonuclear tomography as a complementary tool for precision nuclear structure studies.

Overall, this study lays the ground work for photonuclear tomography as a promising avenue for advancing our understanding of nuclear matter distributions and collective dynamics within atomic nuclei.

\begin{acknowledgments}
C.A.B. acknowledges support  by the ExtreMe Matter Institute EMMI at the GSI Helmholtzzentrum f\"ur Schwerionenforschung, Darmstadt, Germany and by the  U.S. Department of Energy grant DE-FG02-08ER41533.
V.P.G. is grateful to the Mainz Institute of Theoretical Physics (MITP) of the Cluster of Excellence PRISMA+ (Project ID 390831469), for its hospitality and partial support during the completion of this work.  V.P.G. was partially supported by CNPq,  FAPERGS and INCT-FNA (Process No. 464898/2014-5). 
\end{acknowledgments}


\begin{thebibliography}{10}
\expandafter\ifx\csname url\endcsname\relax
  \def\url#1{\texttt{#1}}\fi
\expandafter\ifx\csname urlprefix\endcsname\relax\def\urlprefix{URL }\fi
\expandafter\ifx\csname href\endcsname\relax
  \def\href#1#2{#2} \def\path#1{#1}\fi

\bibitem{ISI:A1988P121800001}
C.~Bertulani, G.~Baur, {Electromagnetic processes in relativistic heavy-ion
  collisions}, {Physics Reports} {163}~({5-6}) ({1988}) {299--408}.
\newblock \href {http://dx.doi.org/10.1016/0370-1573(88)90142-1}
  {\path{doi:10.1016/0370-1573(88)90142-1}}.

\bibitem{annurev.nucl.55.090704.151526}
C.~A. Bertulani, S.~R. Klein, J.~Nystrand, Physics of ultra-peripheral nuclear
  collisions, Annual Review of Nuclear and Particle Science 55~(Volume 55,
  2005) (2005) 271--310.
\newblock \href {http://dx.doi.org/10.1146/annurev.nucl.55.090704.151526}
  {\path{doi:10.1146/annurev.nucl.55.090704.151526}}.

\bibitem{PhysRevC.60.014903}
S.~R. Klein, J.~Nystrand, Exclusive vector meson production in relativistic
  heavy ion collisions, Phys. Rev. C 60 (1999) 014903.
\newblock \href {http://dx.doi.org/10.1103/PhysRevC.60.014903}
  {\path{doi:10.1103/PhysRevC.60.014903}}.

\bibitem{typel:2001:PRC}
S.~Typel, B.~A. Brown, Neutron radii and the neutron equation of state in
  relativistic models, Phys. Rev. C 64~(2) (2001) 027302.
\newblock \href {http://dx.doi.org/10.1103/PhysRevC.64.027302}
  {\path{doi:10.1103/PhysRevC.64.027302}}.

\bibitem{centelles:2009:PRL}
M.~Centelles, X.~Roca-Maza, X.~ViÃ±as, M.~Warda, Nuclear symmetry energy
  probed by neutron skin thickness of nuclei, Phys. Rev. Lett. 102~(12) (2009)
  122502.
\newblock \href {http://dx.doi.org/10.1103/PhysRevLett.102.122502}
  {\path{doi:10.1103/PhysRevLett.102.122502}}.

\bibitem{TamiPRL.107.062502}
A.~Tamii, et~al., Complete electric dipole response and the neutron skin in
  $^{208}\mathrm{Pb}$, Phys. Rev. Lett. 107 (2011) 062502.
\newblock \href {http://dx.doi.org/10.1103/PhysRevLett.107.062502}
  {\path{doi:10.1103/PhysRevLett.107.062502}}.

\bibitem{PhysRevLett.108.112502}
S.~Abrahamyan, et~al., Measurement of the neutron radius of $^{208}\mathrm{Pb}$
  through parity violation in electron scattering, Phys. Rev. Lett. 108 (2012)
  112502.
\newblock \href {http://dx.doi.org/10.1103/PhysRevLett.108.112502}
  {\path{doi:10.1103/PhysRevLett.108.112502}}.

\bibitem{AumannPRL119}
T.~Aumann, C.~A. Bertulani, F.~Schindler, S.~Typel, Peeling off neutron skins
  from neutron-rich nuclei: Constraints on the symmetry energy from
  neutron-removal cross sections, Phys. Rev. Lett. 119 (2017) 262501.
\newblock \href {http://dx.doi.org/10.1103/PhysRevLett.119.262501}
  {\path{doi:10.1103/PhysRevLett.119.262501}}.

\bibitem{Lattimer:2001}
J.~M. Lattimer, M.~Prakash, Neutron star structure and the equation of state,
  The Astrophysical Journal 550~(1) (2001) 426--442.
\newblock \href {http://dx.doi.org/10.1086/319702} {\path{doi:10.1086/319702}}.

\bibitem{2005PhR...411..325S}
A.~W. {Steiner}, M.~{Prakash}, J.~M. {Lattimer}, P.~J. {Ellis}, {Isospin
  asymmetry in nuclei and neutron stars}, Physics Reports 411~(6) (2005)
  325--375.
\newblock \href {http://dx.doi.org/10.1016/j.physrep.2005.02.004}
  {\path{doi:10.1016/j.physrep.2005.02.004}}.

\bibitem{Lattimer:2012}
J.~M. Lattimer, The nuclear equation of state and neutron star masses, Annual
  Review of Nuclear and Particle Science 62~(1) (2012) 485--515.
\newblock \href {http://dx.doi.org/10.1146/annurev-nucl-102711-095018}
  {\path{doi:10.1146/annurev-nucl-102711-095018}}.

\bibitem{PhysRevLett.106.252501}
X.~Roca-Maza, M.~Centelles, X.~Vi\~nas, M.~Warda, Neutron skin of
  $^{208}\mathrm{Pb}$, nuclear symmetry energy, and the parity radius
  experiment, Phys. Rev. Lett. 106 (2011) 252501.
\newblock \href {http://dx.doi.org/10.1103/PhysRevLett.106.252501}
  {\path{doi:10.1103/PhysRevLett.106.252501}}.

\bibitem{HorowitzPRL.86.5647}
C.~J. Horowitz, J.~Piekarewicz, Neutron star structure and the neutron radius
  of $^{208}${Pb}, Phys. Rev. Lett. 86 (2001) 5647--5650.
\newblock \href {http://dx.doi.org/10.1103/PhysRevLett.86.5647}
  {\path{doi:10.1103/PhysRevLett.86.5647}}.

\bibitem{RevModPhys.28.214}
R.~Hofstadter, Electron scattering and nuclear structure, Rev. Mod. Phys. 28
  (1956) 214--254.
\newblock \href {http://dx.doi.org/10.1103/RevModPhys.28.214}
  {\path{doi:10.1103/RevModPhys.28.214}}.

\bibitem{PhysRevA.83.012516}
W.~N\"ortersh\"auser, R.~S\'anchez, G.~Ewald, A.~Dax, J.~Behr, P.~Bricault,
  B.~A. Bushaw, J.~Dilling, M.~Dombsky, G.~W.~F. Drake, S.~G\"otte, H.-J.
  Kluge, T.~K\"uhl, J.~Lassen, C.~D.~P. Levy, K.~Pachucki, M.~Pearson,
  M.~Puchalski, A.~Wojtaszek, Z.-C. Yan, C.~Zimmermann, Isotope-shift
  measurements of stable and short-lived lithium isotopes for
  nuclear-charge-radii determination, Phys. Rev. A 83 (2011) 012516.
\newblock \href {http://dx.doi.org/10.1103/PhysRevA.83.012516}
  {\path{doi:10.1103/PhysRevA.83.012516}}.

\bibitem{PhysRevLett.126.172502}
D.~Adhikari, et~al., Accurate determination of the neutron skin thickness of
  $^{208}\mathrm{Pb}$ through parity-violation in electron scattering, Phys.
  Rev. Lett. 126 (2021) 172502.
\newblock \href {http://dx.doi.org/10.1103/PhysRevLett.126.172502}
  {\path{doi:10.1103/PhysRevLett.126.172502}}.

\bibitem{BertulaniKucuk2025}
C.~A. Bertulani, Y.~Kucuk, F.~Navarra, Nuclear fragmentation at the future
  electron-ion collider, Nuclear Physics A 1059 (2025) 123093.
\newblock \href {http://dx.doi.org/10.1016/j.nuclphysa.2024.123093}
  {\path{doi:10.1016/j.nuclphysa.2024.123093}}.

\bibitem{Sengul:2015ira}
M.~Y. \c{S}eng\"ul, M.~C. G\"u\c{c}l\"u, O.~Mercan, N.~G. Karaku\c{s},
  {Electromagnetic heavy-lepton pair production in relativistic heavy-ion
  collisions}, Eur. Phys. J. C 76~(8) (2016) 428.
\newblock \href {http://dx.doi.org/10.1140/epjc/s10052-016-4269-4}
  {\path{doi:10.1140/epjc/s10052-016-4269-4}}.

\bibitem{Xu:2024dja}
K.~Xu, B.~Chen, {A Study of the Neutron Skin of Nuclei with Dileptons in
  Nuclear Collisions}, Symmetry 16~(9) (2024) 1195.
\newblock \href {http://dx.doi.org/10.3390/sym16091195}
  {\path{doi:10.3390/sym16091195}}.
  
\bibitem{Asche19} E. C. Aschenauer, S. Fazio, J. H. Lee, H. M\"antysaari,
B. S. Page, B. Schenke, T. Ullrich, R. Venugopalan
and P. Zurita, {The electron–ion collider: assessing the
energy dependence of key measurements}, Rept. Prog.
Phys. 82 (2019) 2024301. \newblock \href {http://dx.doi.org/10.1088/1361-6633/aaf216}
  {\path{doi:10.1088/1361-6633/aaf216}}.  
  
\bibitem{Maanfa22}
Heikki M\"antysaari, Farid Salazar, Bj\"orn Schenke,  
{Nuclear geometry at high energy from exclusive vector meson production}. [arXiv:2207.03712] 
\newblock \href {
https://doi.org/10.48550/arXiv.2207.03712}
  {\path{doi:10.48550/arXiv.2207.03712}}.

\bibitem{Maanfa23}
Heikki M\"antysaari, Farid Salazar, Bj\"orn Schenke, Chun Shen, Wenbin Zhao, 
{Effects of nuclear structure and quantum interference on diffractive vector meson production in ultra-peripheral nuclear collisions}. [arXiv:2310.15300]
\newblock \href {https://doi.org/10.48550/arXiv.2310.15300}
  {\path{doi:10.48550/arXiv.2310.15300}}.

\bibitem{1960AnPhy..11....1S}
J.~J. {Sakurai}, {Theory of strong interactions}, Annals of Physics 11~(1)
  (1960) 1--48.
\newblock \href {http://dx.doi.org/10.1016/0003-4916(60)90126-3}
  {\path{doi:10.1016/0003-4916(60)90126-3}}.

\bibitem{PAUL1985203}
E.~Paul, Photoproduction of vector mesons, Nuclear Physics A 446~(1) (1985)
  203--218.
\newblock \href {http://dx.doi.org/10.1016/0375-9474(85)90589-5}
  {\path{doi:10.1016/0375-9474(85)90589-5}}.

\bibitem{SCHULER1993539}
G.~A. Schuler, T.~Sj\"ostrand, {Towards a complete description of high-energy
  photoproduction}, Nuclear Physics B 407~(3) (1993) 539.
  \newblock \href {  https://doi.org/10.1016/0550-3213(93)90091-3}
  {\path{doi:10.1016/0550-3213(93)90091-3}}.


\bibitem{Yin2021-qj}
P.-L. Yin, Z.-F. Cui, C.~D. Roberts, J.~Segovia, Masses of positive- and
  negative-parity hadron ground-states, including those with heavy quarks, Eur.
  Phys. J. C Part. Fields 81~(4).
\newblock \href {http://dx.doi.org/10.1140/epjc/s10052-021-09097-6}
  {\path{doi:10.1140/epjc/s10052-021-09097-6}}.

\bibitem{PhysRevC.65.054905}
V.~P. Gon\ifmmode~\mbox{\c{c}}\else \c{c}\fi{}alves, C.~A. Bertulani,
  Peripheral heavy ion collisions as a probe of the nuclear gluon distribution,
  Phys. Rev. C 65 (2002) 054905.
\newblock \href {http://dx.doi.org/10.1103/PhysRevC.65.054905}
  {\path{doi:10.1103/PhysRevC.65.054905}}.

\bibitem{PhysRevC.71.054902}
R.~Vogt, Shadowing and absorption effects on {J}/$\psi$ production in d-a
  collisions, Phys. Rev. C 71 (2005) 054902.
\newblock \href {http://dx.doi.org/10.1103/PhysRevC.71.054902}
  {\path{doi:10.1103/PhysRevC.71.054902}}.

\bibitem{PhysRevLett.35.483}
U.~Camerini, J.~G. Learned, R.~Prepost, C.~M. Spencer, D.~E. Wiser, W.~W. Ash,
  R.~L. Anderson, D.~M. Ritson, D.~J. Sherden, C.~K. Sinclair, Photoproduction
  of the $\psi$ particles, Phys. Rev. Lett. 35 (1975) 483--486.
\newblock \href {http://dx.doi.org/10.1103/PhysRevLett.35.483}
  {\path{doi:10.1103/PhysRevLett.35.483}}.

\bibitem{PhysRevC.108.025201}
S.~Adhikari, et~al., Measurement of the {J}/$\psi$ photoproduction cross
  section over the full near-threshold kinematic region, Phys. Rev. C 108
  (2023) 025201.
\newblock \href {http://dx.doi.org/10.1103/PhysRevC.108.025201}
  {\path{doi:10.1103/PhysRevC.108.025201}}.

\bibitem{Alexas2013}
C.~Alexa, et~al., Elastic and proton-dissociative photoproduction of {J}/$\psi$
  mesons at {HERA}, The European Physical Journal C 73~(6) (2013) 2466.
\newblock \href {http://dx.doi.org/10.1140/epjc/s10052-013-2466-y}
  {\path{doi:10.1140/epjc/s10052-013-2466-y}}.

\bibitem{Chekanov2002}
S.~Chekanov~et al., T.~Z. Collaboration, Exclusive photoproduction of
  {J}/$\psi$mesons at {HERA}, The European Physical Journal C - Particles and
  Fields 24~(3) (2002) 345--360.
\newblock \href {http://dx.doi.org/10.1007/s10052-002-0953-7}
  {\path{doi:10.1007/s10052-002-0953-7}}.

\bibitem{Acharya2019}
S.~Acharya, , et~al., Energy dependence of exclusive {J}/$\psi$ photoproduction
  off protons in ultra-peripheral p--{Pb} collisions at $\sqrt{s_{NN}} = 5.02$
  {TeV}, The European Physical Journal C 79~(5) (2019) 402.
\newblock \href {http://dx.doi.org/10.1140/epjc/s10052-019-6816-2}
  {\path{doi:10.1140/epjc/s10052-019-6816-2}}.

\bibitem{PhysRevLett.74.1071}
X.~Ji, Qcd analysis of the mass structure of the nucleon, Phys. Rev. Lett. 74
  (1995) 1071--1074.
\newblock \href {http://dx.doi.org/10.1103/PhysRevLett.74.1071}
  {\path{doi:10.1103/PhysRevLett.74.1071}}.

\bibitem{PhysRevLett.89.272302}
C.~Adler, et~al., Coherent ${\ensuremath{\rho}}^{0}$ production in
  ultraperipheral heavy-ion collisions, Phys. Rev. Lett. 89 (2002) 272302.
\newblock \href {http://dx.doi.org/10.1103/PhysRevLett.89.272302}
  {\path{doi:10.1103/PhysRevLett.89.272302}}.

\bibitem{Sirunyan-rho-prod-2019}
A.~Sirunyan, et~al., Measurement of exclusive $\rho^0$(770) photoproduction in
  ultraperipheral ppb collisions at $\sqrt{s_{NN}}=5.02$ {TeV}, The European
  Physical Journal C 79~(8) (2019) 702.
\newblock \href {http://dx.doi.org/10.1140/epjc/s10052-019-7202-9}
  {\path{doi:10.1140/epjc/s10052-019-7202-9}}.


\bibitem{10.1038/ncomms1472}
G.~Aad, et~al., Measurement of the inelastic proton--proton cross-section at
  $\sqrt{s}=7$ {TeV} with the {ATLAS} detector, Nature Communications 2~(1)
  (2011) 463.
\newblock \href {http://dx.doi.org/10.1038/ncomms1472}
  {\path{doi:10.1038/ncomms1472}}.

\end{thebibliography}
\end{document}